\documentstyle[aaspp4,flushrt]{article} \input epsf

\begin{document}
\newcommand\approxgt{\mbox{$^{>}\hspace{-0.24cm}_{\sim}$}}
\newcommand\approxlt{\mbox{$^{<}\hspace{-0.24cm}_{\sim}$}}

\newcommand{\vertsp}{\vphantom{\displaystyle{\dot a \over a}}}
\newcommand{\se}{{(0)}}
\newcommand{\ve}{{(1)}}
\newcommand{\te}{{(2)}}
\newcommand{\nnu}{\nu}
\newcommand{\Spy}[3]{\, {}_{#1}^{\vphantom{#3}} Y_{#2}^{#3}}
\newcommand{\Gm}[3]{\, {}_{#1}^{\vphantom{#3}} G_{#2}^{#3}}
\newcommand{\Spin}[4]{\, {}_{#2}^{\vphantom{#4}} {#1}_{#3}^{#4}}
\newcommand{\scpot}{{\cal V}}
\newcommand{\tl}{\tilde}
\newcommand{\bm}{\boldmath}
\def\bi#1{\hbox{\boldmath{$#1$}}}

\renewcommand{\ell}{l}
\renewcommand{\topfraction}{1.0}
\renewcommand{\bottomfraction}{1.0}
\renewcommand{\textfraction}{0.00}
\renewcommand{\dbltopfraction}{1.0}

\newcommand{\beq}{\begin{equation}}
\newcommand{\eeq}{\end{equation}}
\newcommand{\beqa}{\begin{eqnarray}}
\newcommand{\eeqa}{\end{eqnarray}}

\newcommand{\lexp}{\mathop{\langle}}
\newcommand{\rexp}{\mathop{\rangle}}
\newcommand{\rexpc}{\mathop{\rangle_c}}

\def\d{\delta}
\def\te{\theta}
\def\ds{\delta_s}
\def\dt{\tilde \delta}
\def\dD{\delta_{\rm D}}
\def\del{\nabla}
\def\knl{k_{n\ell}}
\def\pl{{\mathsf P}}
\def\nb{\bar n}

\font\BF=cmmib10
\font\BFs=cmmib10 scaled 833
\def\k{{\hbox{\BF k}}}
\def\x{{\hbox{\BF x}}}
\def\r{{\hbox{\BF r}}}
\def\s{{\hbox{\BF s}}}
\def\ks{{\hbox{\BFs k}}}
\def\xs{{\hbox{\BFs x}}}
\def\q{{\hbox{\BF q}}}
\def\v{{\hbox{\BF v}}}
\def\u{{\hbox{$u_z$}}}
\def\tk{\hat k}
\def\tvk{{\hat{\k}}}
\def\ttheta{\hat \theta}
\def\tphi{\hat \varphi}

\def\la{\mathrel{\mathpalette\fun <}}
\def\ga{\mathrel{\mathpalette\fun >}}
\def\fun#1#2{\lower3.6pt\vbox{\baselineskip0pt\lineskip.9pt
\ialign{$\mathsurround=0pt#1\hfill##\hfil$\crcr#2\crcr\sim\crcr}}}


\title{CMBFAST for spatially closed universes}

\author{Matias Zaldarriaga\footnote{matiasz@ias.edu}} 

\affil{Institute for Advanced Study, School of Natural Sciences,
Olden Lane, Princeton, NJ 08540}

\vskip 1pc

\author{Uros Seljak\footnote{uros@feynman.princeton.edu}}

\affil{Princeton University}


\begin{abstract}

We extend the cosmological linear perturbation theory code 
CMBFAST to closed geometries. This completes the implementation of 
CMBFAST to all types of geometries and allows the user to perform an 
unlimited search in the parameter space of models. This will be 
specially useful for placing confidence limits on cosmological 
parameters from existing and future data. We discuss some of the technical 
issues regarding the implementation. 
\end{abstract}

\keywords{large-scale structure of universe; methods: numerical;
methods: statistical}

\clearpage

\section{Introduction}
\label{intro}

The study of the anisotropies in the Cosmic Microwave Background (CMB)
is revolutionizing cosmology. Fluctuations across a range of 
angular scales
have been detected and their precision in the measurement of the
power spectra are improving steadily. Ground based and balloon born
experiments are making effort to detect the acoustic features in the
power spectrum predicted by inflationary models. In the near future the MAP
satellite and a few years later the Planck satellite 
will be able to measure
the CMB power spectrum with sufficient accuracy to constraint several
cosmological parameters with a few percent accuracy
percent accuracy (\cite{jungman,bond,zss}).

There are two reasons that explain why the CMB is so powerful 
at constraining the cosmological parameters. First is
the improvement in the experimental sensitivity that allows the detection of 
these small anisotropies with increasing precision. Second is our
ability to accurately calculate the predictions for the different world
models. These theoretical predictions can be made with high accuracy
because the calculation of the power spectrum 
only involves linear perturbation theory. The same fact that makes the
anisotropies difficult to detect makes them also easy to calculate. 

The most widely used Boltzmann code to calculate anisotropy
spectra, CMBFAST, was developed by the authors and has become a
standard tool freely 
available to the community \cite{cmbfast1,cmbfast2}. The code is based
on the integral solution of the transport equation for the photons,
the Boltzmann equation.  
The algorithm is much more efficient than the traditional way of
solving the hierarchy because it takes advantage of the large
difference between the distance photons have traveled since
recombination and the horizon at that time.

There have been several studies intended to constrain cosmological
parameters with the existing CMB data \cite{lineweaver,maxpar,lksd}. These
studies rely on the speed of CMBFAST to search a
multidimensional parameter space for the best fitting model and the
confidence regions for the parameters. They typically evaluate millions
of models which would be prohibitely slow with a traditional Boltzmann code
taking a hundred 
times longer than CMBFAST to run. So far the 
available versions of CMBFAST did not allow the user to explore models
with spatially closed geometries and thus that part of parameter space
has not been explored with this code (although  
closed universe Boltzmann codes do exist, see e.g. \cite{scottwhite}). 
Since the data are pushing us into the corner
of parameter space with vanishing curvature, any meaningful 
error determination must also explore the models with positive curvature.
It is therefore timely to make CMBFAST available also for this class
of models.

A key part of the CMBFAST algorithm is being able to calculate
efficiently the radial part of the eigenfunctions of the 
Laplacian, the ultraspherical
Bessel functions. In our implementation of open models we calculate
these functions by integrating their corresponding differential
equation. The straightforward extension of this algorithm is not 
possible for closed models. In this paper we explain the technical
difficulties that arise for closed models and present the solution
that is implemented in the new version of
CMBFAST (V3.0). 
 
\section{The line of sight method and a
brief description of geometries}

In spherical coordinates the unperturbed geometry is described by,
\begin{equation}
ds^2=a^2(\tau) [- d \tau^2+d\chi^2+r^2(\chi)(d\theta^2+\sin^2 \theta d\phi^2)].
\label{metric}
\end{equation}
The scale factor is $a(\tau)$, the coordinate $\tau$ is conformal time
and the radial 
coordinate $\chi$ is related to the angular radius $r$ by
\begin{eqnarray}
r(\chi)\equiv
\left\{ \begin{array}{ll} K^{-1/2}\sin K^{1/2}\chi,\ K>0\\
\chi, \ K=0\\
(-K)^{-1/2}\sinh (-K)^{1/2}\chi,\ K<0.\\
\end{array}
\right.
\label{rchi}
\end{eqnarray}
with $K=H^2_0(\Omega_0-1)$, where $\Omega_0$ includes the contribution
from both matter and the cosmological constant. Note that with our
choice of variables  we have $d\tau=-d\chi$ along radial geodesics.

To calculate the CMB power spectrum $C_l$ we need to evolve the
perturbations in the different species (CDM, baryons, photons,
neutrinos). The perturbations are so small that 
linear theory suffices. The standard
method is first to expand all perturbations in eigenvalues of the
Laplacian which evolve independently. In a flat spacetime this is a 
Fourier expansion. The wavenumber $\beta$ labels the different
eigenmodes. The fluctuations produced by each mode can be calculated
separately and then all the contributions to the power are added,
\begin{equation}\label{Cls}
C_l=(4\pi)^2\int \beta d\beta \ P(\beta) |\Delta_{Tl}(\beta)|^2.
\end{equation}
We have introduced $\Delta_{Tl}(\beta)$, the contribution by a mode
of wavenumber $\beta$ to the amplitude of multipole $l$ of the
temperature anisotropies. The
primordial power spectrum is denoted by $P(\beta)$. Equation
(\ref{Cls}) says that to obtain the total power in multipole $l$
we add the contributions from all modes labeled by $\beta$. We have
written this sum as a continuous integral, which applies to flat and
open models. For closed spatial hypersurfaces the eigenvalues of the
Laplacian are discrete so the integral becomes
a discrete sum over $\beta$ with
$K^{-1/2}\beta=3,4,5,\cdots$. 
The first two modes are pure gauge modes
and should be excluded from the sum \cite{abbott}. 

In CMBFAST the amplitudes $\Delta_{Tl}$ are obtained using the
integral solution,
\begin{equation}\label{integsol}
\Delta_{Tl}=\int_0^{\tau_0}d\tau
\Phi_\beta^l(\chi)S(\beta,\tau). 
\end{equation}
The ultraspherical Bessel functions $\Phi_\beta^l(\chi)$ 
depend on the wavenumber $\beta$ and the radial coordinate along the
geodesic $\chi=\tau_0-\tau$.  The source $S(\beta,\tau)$ is a sum of the different
terms that can generate temperature anisotropies. Among these are
gravitational
redshift, Doppler shift and intrinsic density fluctuation in the photon
baryon fluid. The dominant contribution to the source is
produced at the last scattering surface. 
The full expression for $S$ together with the
equivalent formulas for the CMB polarization can be found in 
\cite{cmbfast2}.

Equation (\ref{integsol}) has several advantages over the more
traditional way of calculating the anisotropies. 
The main one is that the sampling in $\beta$ needed to compute
accurately the source $S$ (essentially the inverse of the sound
horizon at recombination) is much smaller than that needed to 
obtain an accurate representation of $\Phi_\beta^l$ (approximately 
the inverse of
the distance to the last scattering surface). In the traditional
Boltzmann method one solves directly for $\Delta_{Tl}$ so one has to
sample $\beta$ with the inverse of the distance to the last scattering
surface. Obtaining the source for
each $\beta$ involves solving the perturbation equations for the
different species so a reduction in the number of equations to be
solved greatly enhances the performance of the code. 

In essence traditional Boltzmann codes spend most of their time
computing Bessel functions. If we had a fast method for computing the
Bessel functions we could use equation (\ref{integsol}) to make a fast
algorithm. This is what CMBFAST does. In flat geometries the
ultraspherical Bessel functions reduce to ordinary spherical Bessel
functions $j_l(\beta\chi)$ which are a function of a single variable
$x=\beta\chi$. In the implementation of CMBFAST for flat models 
we use  tables with the values of the spherical Bessel functions which
are generated in advance and stored in disk. 

For open model we are unable to tabulate the ultraspherical Bessel
functions in advance because they no longer depend on only one
variable. They depend separately on $\beta$ and $\chi$ because 
the curvature radius of the universe introduces another scale that
can be used to turn $\beta$ and
$\chi$ into two dimensionless numbers and the ultraspherical
Bessel functions depend on these two numbers independently. In our
implementation of CMBFAST for open models we choose to calculate 
the Bessel function by solving their differential equation, 
\begin{equation}\label{diffeq}
{d^2 u_{\beta}^l\over d\chi^2}+\left[\beta^2-{l(l+1)\over r(\chi)^2}\right]
u_{\beta}^l=0
\end{equation}
where $u_{\beta}^l(\chi)=r(\chi)\Phi_{\beta}^l(\chi)$.

Although at first this may not seem to be the most efficient method
(for example, 
we could have used the recurrence relation for the ultraspherical 
Bessel functions
instead), it is actually very efficient
for our purposes. The point is that we need the function at
a succession of neighboring points in time when evaluating
equation (\ref{integsol}). To
get from one point to the next a Runge Kutta integrator only performs a
handful of operations, as opposed to the recurrence relation calculation
which has to start at sufficiently high $l$. It is true that recurrence
relation gives the function value at all $l$, but since the final CMB 
spectra are so smooth we actually only use every 50th $l$ or so, so 
this advantage of recurrence relation becomes less significant.
Thus although the recurrence relation is the
most efficient way to obtain a single value of the Bessel function,
for our purposes solving the differential equation is a better
method than the recurrence relation. 
Another method to calculate these Bessel functions 
based on the WKB approximation has been proposed
(\cite{arthur}). This method is again a 
fast way to obtain a single value
of any $\Phi_\beta^l$ since 
it only involves a few operations per  $\Phi_\beta^l$
call. However, for the integral solution which requires many
calls at nearby arguments, the available implementation of these formulas
does not appear to be faster than solving the differential
equations. 

\section{Closed models}

In \cite{cmbfast2} we had developed the necessary formalism to solve
the perturbation equations in both open and closed universes. We
derived the Boltzmann hierarchy for both temperature and polarization
and the integral solution. The other trivial differences between
closed and open models like the discreteness of the eigenvalues of the
Laplacian were also mentioned. The only problem not addressed was the
calculation of the Bessel functions in closed models. We discuss 
this issue here.

Equation (\ref{diffeq}) is a second order differential equation and
thus has two independent solutions. It 
can be viewed as the Schr\" odinger equation 
for a particle with
energy $E=\beta^2$ in a potential 
$V=l(l+1) / r(\chi)^2$. Thus for $\chi$ such that
$\beta  r(\chi) > \sqrt{l(l+1)}$, where the ``energy''
is greater than the ``potential'', the solution is oscillatory.
For $\beta r(\chi) < \sqrt{l(l+1)}$ 
there is a growing and a decaying solution, $u_{\beta}^l\propto r^{l+1}$
and $u_{\beta}^l\propto r^{-l}$, respectively.
The growing solution corresponds to $\Phi_{\beta}^l(\chi)$. In order to 
obtain 
an  accurate numerical convergence the integration needs to be started 
in the regime where $\Phi_{\beta}^l(\chi)$ is small, 
which in our case means starting
the integration close to $\beta r(\chi)\approx l$. The equation
then needs to be evolved in the direction of increasing $\chi$  
until recombination, where $\chi \sim \tau_0$. 
The integration of $\Phi_{\beta}^l(\chi)$
therefore proceeds in the direction of decreasing time, opposite 
to the evolution of Boltzmann, fluid and Einstein equations. 
If one were to start the integration at early time and evolve
the Bessel functions to small radial distances $\chi$ (i.e. 
present time)
the integration would be numerically
unstable, as the decaying mode $u_{\beta}^l\propto r^{-l}$ 
would increasingly contaminate the solution.   

\begin{figure*}
\begin{center}
\leavevmode
\epsfxsize=4.0in \epsfbox{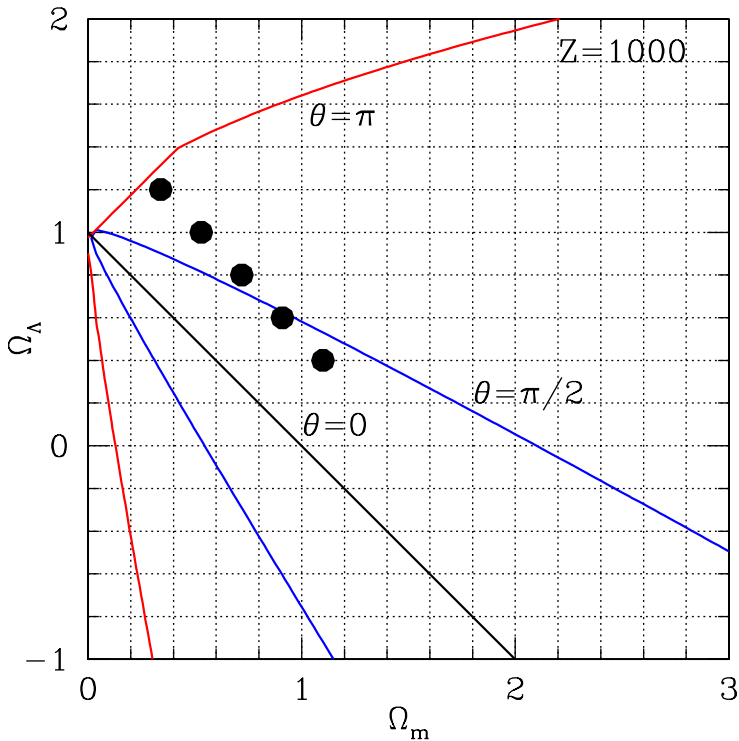}
\end{center}
\caption{Contours of constant $\theta$ in the
$(\Omega_m,\Omega_\Lambda)$ plane. The line $\theta=0$ corresponds to
flat models, with closed models above that line. In the region 
between $\theta=\pi/2$ and $\theta=\pi$ the
calculation of the Bessel functions becomes unstable and symmetry can 
be used to determine their values in that regime. The dots show
the positions of the models we show in figure \ref{fig2}. The two
bottom lines also correspond to $\theta=\pi/2,\pi$ but in an open
universe.}
\label{fig1}
\end{figure*}

In closed models the angular distance $r(\chi)$ is periodic
around $\theta \equiv K^{-1/2}\chi=\pi/2$. The potential
$V(\chi)$ is a potential well.
For parameters such that the maximum value of $\theta$ required in the
evaluation of equation (\ref{integsol}) is greater than $\pi/2$ 
the integration of the differential equation (\ref{diffeq}) 
becomes unstable, similarly to integrating in the direction of
decreasing $\chi$ in the 
open model case. As the integration proceeds
the solution becomes increasingly contamined by the decaying solution which
diverges at $\theta=\pi$. In figure \ref{fig1} we show lines of constant
$\theta$ in the $(\Omega_m,\Omega_\Lambda)$ plane for $z_{max}=1000$. 
The region upwards of the $\theta=\pi/2$ line suffers from this instability.

We can use symmetry arguments analogous to those in 
quantum mechanics to solve the
instability 
problem. The potential well $V(\chi)$ is symmetric around
$\theta=\pi/2$ and thus the eigenfunctions are either symmetric or
antisymmetric around this point. Just like in quantum mechanics we
find that modes with even (odd) $\beta - l -1$ 
are symmetric (antisymmetric). In our implementation of CMBFAST
for closed models we use this fact to obtain the Bessel functions for
regions with $\theta>\pi/2$. We integrate equation (\ref{diffeq}) up to
$\theta=\pi/2$ and extend the solution further using the symmetry of
the mode. In practice we apply the symmetry operation to  
the source term $S(\beta,\tau)$. We construct a
symmetric $S^s=(S(\beta,\tau)+S(\beta,\tau^\prime))/2$ and an
antisymmetric $S^a=(S(\beta,\tau)-S(\beta,\tau^\prime))/2$ version of the
source. Here $\tau$ and $\tau^{\prime}$ are times that 
have the same $r(\chi)$, related by $\tau^{\prime}=K^{1/2}\pi-\tau$.
We use one of those in (\ref{integsol}) according to the parity
of the mode being calculated and stop the time integration at
the $\tau_{\pi/2}=\tau_0-K^{1/2}\pi/2$ corresponding to 
$\theta=\pi/2$. 
We found this to be both a  satisfactory solution of the
instability and an efficient numerical algorithm.

\begin{figure*}
\begin{center}
\leavevmode
\epsfxsize=4.0in \epsfbox{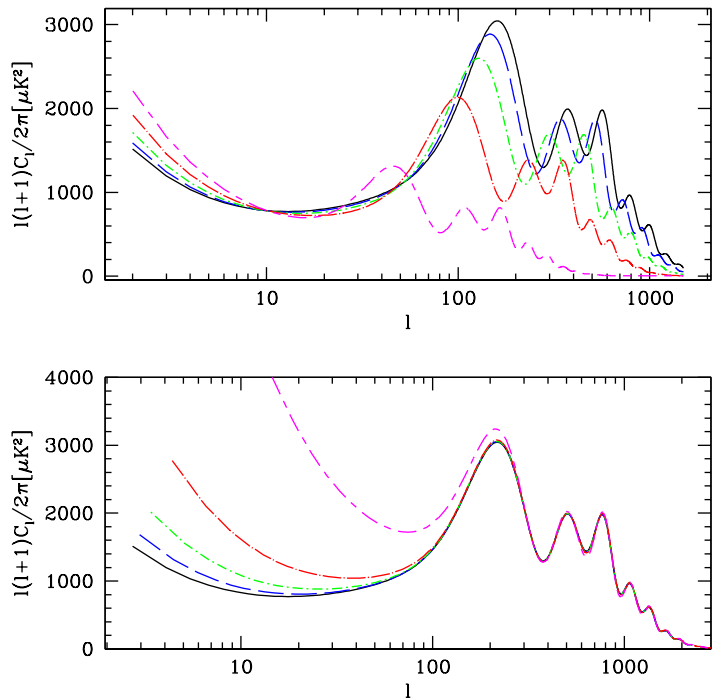}
\end{center}
\caption{Sample of closed universe models calculated with the new version of
CMBFAST. Panel b shows the same curves but rescaled in $l$ and
amplitude to match as closely as possible.}
\label{fig2}
\end{figure*}

In figure \ref{fig2}a we show the power spectra obtained using the new
CMBFAST version for several closed models. We choose models with
$\Omega_m h^2=0.237$, 
$\Omega_b h^2=0.013$ and varying
$\Omega_{\Lambda}=0.4,0.6,0.8,1.0,1.2$.
The position of the models in the
$(\Omega_m,\Omega_\Lambda)$ plane is shown in figure \ref{fig1}.
The models we show all have the same
values of $\Omega_m h^2$ and $\Omega_b h^2$, so the physics of the
acoustic peaks is the same for all the models. They only differ in the
angular diameter distance to the last scattering surface and their
amplitude which was normalized to COBE.
In figure \ref{fig2}b we
rescale the $l$ values so that a given physical scale at recombination
corresponds 
to the same angular scale. This is achieved using $l
\rightarrow l
r^{\prime}(\chi_R)/r(\chi_R^{\prime})$, where $r(\chi_R)$ is the 
radial comoving distance to recombination  and primed variables
correspond to the values in a fiducial model. 
We also renormalized the amplitude of the spectra
so they are no longer COBE normalized. 
The agreement at high $l$ overall is very good, which is a
consistency check for the code.

As we approach $\theta=\pi$, $r(\chi)\rightarrow 0$, when we are  
able to see the antipode. As $\theta \rightarrow \pi$ the $C_l$
spectra is compressed so that the acoustic peaks appear at larger
and larger scales. When $\theta=\pi$ all the CMB photons that arrive
to us from every direction come form one single point, so there are no
anisotropies. We have restricted the range of models implemented in
CMBFAST to those with $\theta < \pi$.

In our
previous implementations of CMBFAST we took advantage of the
smoothness of the $C_l$ spectra to calculate (\ref{integsol}) for
every 50th $l$ in the
region of the acoustic peaks. This was sufficient to obtain one percent
accuracy in flat models and became even better for open model were the
spectra are stretched further out in $l$. 
For closed models the situation is reversed because the spectra are
compressed so this sampling can 
become insufficient. If the first acoustic peak occurs at $l\sim 130 (100)$
as opposed to $l\sim 200$ for flat universes, the sampling every 50
ls introduces an error of 2\% (5\%).

\section{Conclusions}

We extended the CMBFAST code to the
closed universes. The main technical issue in this implementation 
was the computation of 
ultraspherical Bessel functions, which was obtained by solving
their differential equation jut like in the case of open models. 
We use the
symmetry of the ultraspherical Bessel   
functions around $\theta=\pi/2$ to obtain their values at  
$\theta>\pi/2$ without integration.
The code only works for models with $\theta<\pi$.
The new version of CMBFAST for closed universes will allow the exploration
of the full parameter space. The results of such an exploration will be
presented elsewhere \cite{tegzal}.

\smallskip
We thank Anthony Challinor and Antony Lewis 
for helpful comparisons with their
version of CMBFAST for closed universes.
M.Z. is supported by NASA through Hubble Fellowship grant
HF-01116.01-98A from STScI,
operated by AURA, Inc. under NASA contract NAS5-26555.
U.S. is supported by NASA grant NAG5-8084.

%
%


\end{document}